\documentclass[useAMS,usenatbib,onecolumn]{mn2e}
\usepackage{epsfig,amssymb,amsmath,graphicx,enumitem}

\newcommand{\be}{\begin{equation}}
\newcommand{\ee}{\end{equation}}
\newcommand{\er}{{\bf{\hat{e}}}_r}

\newcommand{\etht}{{\bf{\hat{e}}}_\theta}

\newcommand{\rstar}{R_\ast}
\newcommand{\mstar}{M_*}

\newcommand{\msun}{M_\odot}
\newcommand{\alphac}{\alpha_{\mathrm{c}}}

\title[Stellar deformation due to multipolar magnetic fields]{Neutron star deformation due to poloidal-toroidal magnetic fields of arbitrary multipole order: a new analytic approach}
\author[A. Mastrano, A. G. Suvorov, and A. Melatos]{A. Mastrano\thanks{E-mail:
alpham@unimelb.edu.au}, A. G. Suvorov\thanks{E-mail: suvorova@student.unimelb.edu.au}, and A.
Melatos\thanks{E-mail: amelatos@unimelb.edu.au}\\School of Physics, University of Melbourne, Parkville VIC
3010, Australia}

\begin{document}

\date{Accepted ?. Received ?; in original form ?}

\pagerange{\pageref{firstpage}--\pageref{lastpage}} \pubyear{?}

\maketitle

\label{firstpage}

\begin{abstract}

\noindent{A recipe is presented to construct an analytic, self-consistent model of a non-barotropic neutron star with a poloidal-toroidal field of arbitrary multipole order, whose toroidal component is confined in a torus around the neutral curve inside the star, as in numerical simulations of twisted tori. The recipe takes advantage of magnetic-field-aligned coordinates to ensure continuity of the mass density at the surface of the torus. The density perturbation and ellipticity of such a star are calculated in general and for the special case of a mixed dipole-quadrupole field as a worked example. The calculation generalises previous work restricted to dipolar, poloidal-toroidal and multipolar, poloidal-only configurations. The results are applied, as an example, to magnetars whose observations (e.g., spectral features and pulse modulation) indicate that the internal magnetic fields may be at least one order of magnitude stronger than the external fields, as inferred from their spin downs, and are not purely dipolar.}


\end{abstract}


\section{Introduction}

Neutron stars, particularly the subset known as magnetars, possess some of the strongest known, naturally occurring magnetic fields. The external field strength of a neutron star is inferred from its spin down. The \emph{internal} field cannot be measured directly. Observations of bursts and giant flares \citep{i01,co11} and precession \citep{metal14} in magnetars have been interpreted to indicate that the internal field exceeds the external field by at least one order of magnitude. Numerical simulations favour a `twisted torus' magnetic configuration \citep{bn06,bs06}.

Magnetic stresses deform a neutron star \citep{cf53,f54,g72,k89,c02,pm04,hetal08,metal11}, leading to detectable levels of gravitational radiation under certain conditions \citep{bg96,mp05,setal05,hetal08,dss09}. Upper limits from gravitational wave non-detections to date can be used to set upper limits on the star's ellipticity, and hence to constrain the strength and topology of the internal field in principle \citep{c02,dss09,metal11,mm12}.

Neutron star magnetic fields are approximately dipolar at radio emission altitudes \citep{lm88,cm11,bm14} and in the outer magnetosphere, where high-energy emissions originate \citep{ry95,lom12}. However, some observations can be interpreted as signatures of higher-order multipoles close to the surface. Examples include intermittent radio emission from pulsars beyond the pair-cascade `death line' \citep{ymj99,cetal00,gm01,metal03,ml10}, the pulse profile of SGR 1900+14 following its 1998 August 27 giant flare \citep{fetal01,td01,tlk02}, cyclotron resonant scattering line energies of some accretion-powered X-ray pulsars \citep{n05,pml14}, the anomalous braking index of some radio pulsars \citep{bt10}, and X-ray spectral features of SGR 0418+5729 \citep{gog08,ggo11,tetal13} and PSR J0821$-$4300 \citep{gha13} . While the dipole component of the magnetic field is readily inferred from its spin down, the putative higher-order multipoles do not contribute much to the spin-down rate.

In analytically studying higher-order multipoles through (say) their gravitational wave emission, one confronts a key technical challenge, namely ensuring that the magnetically-induced pressure and density perturbations are continuous everywhere. If the field has both poloidal and toroidal components, and the form of the toroidal component is not chosen wisely, one ends up with discontinuous pressure and density profiles. The `correct' form of the toroidal field (i.e., one that is analytically tractable and does not lead to discontinuities) is not immediately obvious. In previous papers, we investigated the relation between field configuration and stellar ellipticity in a stratified, non-barotropic star with an axisymmetric, purely dipolar, poloidal and toroidal field \citep{metal11} and an axisymmetric, multipolar, purely poloidal field \citep{mlm13}. Realistic neutron stars are non-barotropic; they are stably stratified by a composition gradient \citep{p92,rg92,r09}. Because a non-barotropic star allows arbitrary relative poloidal and toroidal field strengths, it can easily accommodate the strong internal toroidal fields suggested by observations \citep{i01,co11,metal14} and simulations \citep{bn06}. However, like other authors, we were unable to construct a physically and mathematically consistent poloidal-toroidal multipolar field. We found that the simple forms of an axisymmetric field used by \citet{metal11} and others lead to unphysical density discontinuities at the boundary of the toroidal field region. There appears to be no record in the literature of an analytic poloidal-toroidal construction for multipoles of higher order than dipole. This is unfortunate, because once this issue is dealt with, linear combinations of multipoles can generate realistic analytic models which complement numerical simulations helpfully.

In this short methods paper, we present a new method for constructing such a consistent axisymmetric poloidal-toroidal field analytically for any multipole. In Sec. 2, we recap the results of \citet{metal11} and \citet{mlm13}, and we describe the method for calculating $\epsilon$ for a generalised multipolar field. In the Appendix, we show why the new method presented here is necessary, by way of an explicit example, where the old methods fail. In Sec. 3, we present a worked example of a mixed dipole-quadrupole, poloidal-toroidal field, constructed with the new method. We give explicit expressions for the field and calculate the deformation due to this field. In Sec. 4, we apply the results briefly to the magnetars SGR 0418+5729 and 4U 0142+61 to give a flavour of the sort of astrophysical questions that the method can help to study. Lastly, in Sec. 5, we summarize our findings. We emphasize that our analytic field models should complement, rather than supplant, large-scale numerical simulations. They are not meant to be true representations of neutron star fields but rather to provide helpful toy models, which are easy to manipulate and whose general properties are easy to parametrize and understand.

\section{Mass ellipticity}

In this section, we describe a general method for constructing an analytic model of a non-barotropic star in hydromagnetic equilibrium, where the poloidal component of the magnetic field is a linear combination of multipolar fields, and the toroidal component is located around the neutral curve (the circumstellar curve inside the star where the poloidal component vanishes). This method is needed to ensure continuity of the density field at the boundary of the magnetic torus; an example of what can go wrong without it is given in the Appendix. The new feature of the method is a field-aligned coordinate transformation which allows one to solve the force balance equation exactly in closed form for the density perturbations, generated by a poloidal magnetic stream function of arbitrary multipole order. Previous calculations \citep{metal11,mm12,y13,detal14} are restricted to dipolar poloidal fields.

\subsection{Hydromagnetic force balance}

We decompose the magnetic field into its poloidal and toroidal components and express it in dimensionless spherical polar coordinates ($r,\theta,\phi$) in the usual way \citep{c56,metal11,mm12,mlm13}, viz.

\be {\bf{B}}=B_0 [\eta_p \nabla\alpha(r,\theta)\times \nabla\phi + \eta_t \beta(\alpha)\nabla\phi],\ee
where $B_0$ parametrizes the overall strength of the field, $\eta_{p}$ and $\eta_t$ set the relative strengths of the poloidal and toroidal components respectively ($\eta_p= 1$ without loss of generality), $\alpha(r,\theta)$ is the poloidal magnetic stream function, and the function $\beta(\alpha)$ defines the toroidal field component. In general, the stream function $\alpha(r,\theta)$ can be expanded as a linear combination of multipolar stream functions $\alpha_l(r,\theta)$ up to any desired order $n$, viz.

\be\alpha(r,\theta)=\sum_{l=1}^n a_l\alpha_l(r,\theta),\ee
where the subscripts denote the multipole orders and $a_l$ is a dimensionless weight. We consider separable stream functions of the form $\alpha_l(r,\theta)=f_l(r)g_l(\theta)$ in this paper. The function $\beta$ must be a function of $\alpha$ to ensure that the magnetic force has no azimuthal component, which cannot be balanced in magnetohydrostatic equilibrium given a field of the form (1) \citep{metal11,mlm13}.

The magnetic energy density is $\lesssim 10^{-6}$ of the gravitational energy density, even in magnetars. Therefore, we can treat the magnetic force as a perturbation on a background hydrostatic equilibrium and write the hydromagnetic force balance equation as

\be\frac{1}{\mu_0} (\nabla\times {\bf{B}})\times{\bf{B}}=\nabla\delta p +\delta\rho\nabla\Phi,\ee
to first order in $B^2/(\mu_0 p)$ in the Cowling approximation $(\delta\Phi = 0)$, where $p$ is the zeroth-order pressure, $\rho$ is the zeroth-order density, $\Phi$ is the gravitational potential, and $\delta p$, $\delta\rho$, $\delta\Phi$ are perturbations of the latter three quantities. Because we do not assume a barotropic star, the density perturbation $\delta\rho$ does not have to be a function solely of the pressure perturbation $\delta p$, and therefore the equation of state imposes no restrictions on the field structure. Physically, this means that the imposed magnetic field sets the density and pressure perturbations, but the resulting perturbations do not restrict the magnetic field in turn. Therefore, unlike some previous works [e.g., \citet{hetal08,lj09,cfg10}], we do not specify a barotropic equation of state and then solve the Grad-Shafranov equation for the magnetic field configuration. Instead, we specify the magnetic field whose effects we wish to investigate [to be determined from observations, e.g., neutron star spin down, gravitational wave upper limits, and radio polarization \citep{cm11,bm14}] then calculate the density perturbations that the field causes.

We characterize the magnetic deformation of the star by its ellipticity $\epsilon$,

\be \epsilon = \frac{I_{zz}-I_{xx}}{I_0},\ee
where $I_0$ is the moment of inertia of the unperturbed spherical star, the moment-of-inertia tensor is given by

\be I_{jk} = R^5_* \int_V \textrm{d}^3x[\rho(r)+\delta\rho(r,\theta)](r^2\delta_{jk}-x_j x_k),\ee
$R_*$ is the stellar radius, and the integral is taken over the volume of the star $(r\leqslant 1)$. The density perturbation $\delta\rho$ appearing in equation (5) is calculated by taking the curl of both sides of equation (3) and matching the $\phi$-components:

\be \frac{\partial\delta\rho}{\partial\theta}=-\frac{r}{\mu_0 \rstar}\frac{\mathrm{d}r}{\mathrm{d}\Phi}\{\nabla\times[(\nabla\times{\bf{B}}\times{\bf{B}})]\}_\phi.\ee
Equations (5) and (6) are then solved to obtain $\epsilon$.

We require the field to obey the following conditions:

\begin{enumerate}[leftmargin=*]
\item the field is symmetric about the $z$-axis;
\item the external field is current-free and is purely poloidal;
\item the poloidal component of the field is continuous everywhere;
\item the toroidal component of the field is confined to a toroidal region inside the star around the neutral curve;
\item the current is finite and continuous everywhere inside the star and vanishes at the surface of the star.
\end{enumerate}
These conditions are to be fulfilled by judicious choices of $\alpha$ and $\beta$.

Formally, the function $\beta(\alpha)$ is arbitrary. In previous work \citep{r09,metal11,aetal13,mlm13}, we take

\be \beta(\alpha) =
\begin{cases}
(|\alpha| - \alpha_c)^2&\textrm{for }|\alpha|\geqslant \alpha_c,\\
0&\textrm{for }|\alpha| < \alpha_c,
\end{cases}
\ee
where $\alphac$ is the value of $\alpha$ that defines the last poloidal field line that closes inside the star. This relatively simple form of $\beta$ ensures that the toroidal field is confined to the equatorial torus bounded by the last closed poloidal field line. However, if the field line $\alpha(r,\theta)=\alphac$ is not symmetric around some $\theta$, this simple polynomial form for $\beta$ leads to a discontinuity in $\delta\rho$ at the $\alpha=\alphac$ boundary \citep{mlm13}. A worked example showing this explicitly for a quadrupolar field is presented in the Appendix. For this reason, previous papers failed to construct multipolar fields with poloidal and toroidal components analytically \citep{metal11,mlm13}.

\subsection{Non-barotropicity}

As noted above, we consider the star to be non-barotropic in this paper, i.e., we do not require a one-to-one relation between pressure and density, $p = p(\rho)$. Neutron star matter consists of multiple species (at least neutrons, protons, and electrons) which reach a stably-stratified, hydromagnetic equilibrium within a few Alfv\'{e}n time-scales \citep{p92,rg92,r01,r13}. This system is not in full chemical equilibrium: the relative abundances of the constituent particles change by weak nuclear interactions and diffusive processes. These processes have much longer time-scales than the Alfv\'{e}n time-scale \citep{hrv08}, with

\be \tau_\textrm{weak} = 4.3\times 10^5 \left(\frac{T}{10^8\textrm{ K}}\right)^{-6} \textrm{ yr},\ee
\be \tau_\textrm{diff} = 1.7\times 10^9 \left(\frac{B}{10^{11}\textrm{ T}}\right)^{-2}\left(\frac{T}{10^8\textrm{ K}}\right)^{-6}\textrm{ yr},\ee
respectively, where $T$ is the temperature of the star. Between the Alfv\'{e}n and the weak nuclear time-scales, the star is in a hydromagnetic equilibrium state, in which the composition is not determined solely by the density or pressure, and density and pressure do not correspond one-to-one \citep{metal11}. Since most magnetars are $\sim 1$--10 kyr old (as inferred from spin and spin down), the calculations presented in this paper are readily applicable to them. In addition, note the strong inverse dependence on $T$; neutron stars cool down rapidly over $\sim 10^2$ kyr \citep{yls99,pr10,pr11,gjpr14}, so that, in practice, chemical equilibrium may never be reached.

\subsection{Field-aligned coordinates}

In this section, we describe a method of generating $\alpha$ that does not lead to discontinuities in $\delta\rho$ while keeping equation (7) for $\beta(\alpha)$. There are no cross terms between the poloidal and toroidal field components in the Lorentz force, so we can calculate the density perturbations $\delta\rho_t$ and $\delta\rho_p$ caused by the toroidal and poloidal field components separately and add them to arrive at the total density perturbation $\delta\rho$. Substituting equation (1) into equation (6), we find

\be -\frac{\mu_0\rstar}{B_0^2}\frac{\mathrm{d}\Phi}{\mathrm{d}r}\frac{\partial\delta\rho_t}{\partial\theta}= \frac{\partial}{\partial\theta}\left(F\beta\beta'\frac{\partial\alpha}{\partial r}\right)-\frac{\partial}{\partial r}\left(F\beta\beta'\frac{\partial\alpha}{\partial\theta}\right)\ee
and

\be
\begin{split}
-\frac{\mu_0\rstar}{B_0^2}\frac{\mathrm{d}\Phi}{\mathrm{d}r}\frac{\partial\delta\rho_p}{\partial\theta} &= \frac{\partial}{\partial\theta}\left\{\frac{1}{r^2\sin\theta} \frac{\partial\alpha}{\partial r} \left[\frac{1}{\sin\theta} \frac{\partial^2\alpha}{\partial r^2} + \frac{1}{r^2} \frac{\partial}{\partial\theta} \left(\frac{1}{\sin\theta} \frac{\partial\alpha}{\partial\theta}\right)\right]\right\}\\
& \quad -\frac{\partial}{\partial r}\left\{\frac{1}{r^2\sin\theta} \frac{\partial\alpha}{\partial \theta} \left[\frac{1}{\sin\theta} \frac{\partial^2\alpha}{\partial r^2} + \frac{1}{r^2} \frac{\partial}{\partial\theta} \left(\frac{1}{\sin\theta} \frac{\partial\alpha}{\partial\theta}\right)\right]\right\},
\end{split}
\ee
for the two components, with $F=(r\sin\theta)^{-2}$ and $\beta'=\mathrm{d}\beta/\mathrm{d}\alpha$.

There are many valid ways to choose $\beta(\alpha)$. However, numerical simulations \citep{bn06,b09} favour configurations where the toroidal field is confined fully inside the star in a closed region around the neutral curve. We therefore seek a function $\alpha$ that, when substituted into $\beta(\alpha)$, ensures continuity of $\delta\rho_t$ at $\alpha=\alphac$. (Note that $\delta\rho_p$ is continuous everywhere, since $\alpha$ is continuous everywhere). This task is made more tractable by first performing a coordinate transformation from spherical polar coordinates to those defined by the stream function $\alpha$ and some coordinate $\gamma$ (which indicates angular position on the meridional field line defined by the coordinate $\alpha$), i.e., $(r,\theta,\phi) \mapsto  (\alpha,\gamma,\phi)$. The coordinates $\alpha$ and $\gamma$ must be chosen such that the Jacobian of this transformation is nonzero in the regions of interest and the transformation is invertible \citep{c11}. Furthermore, if $\gamma$ is chosen to satisfy $\partial\gamma/\partial r = 0$ everywhere, equation (10) can be rewritten as

\be - \frac{\partial\alpha}{\partial\theta}\frac{\partial\delta\overline{\rho}_t(\alpha,\gamma)}{\partial\alpha} - \frac{\partial\gamma}{\partial\theta}\frac{\partial\delta\overline{\rho}_t(\alpha,\gamma)}{\partial\gamma} = \frac{\partial F(\alpha,\gamma)}{\partial\gamma}\beta(\alpha)\beta'(\alpha)\frac{\partial\alpha}{\partial r}\frac{\partial\gamma}{\partial\theta},\ee
with $\delta\overline{\rho}_t = \mu_0\rstar \delta\rho_t \mathrm{d}\Phi/\mathrm{d}r$. If $\alpha$ and $\gamma$ are chosen such that $\partial\alpha/\partial\theta$, $\partial\gamma/\partial\theta$, and the entire right-hand side of equation (12) are all functions of $\alpha$ and $\gamma$ only, then equation (12) is a well-defined, first-order, linear, inhomogeneous partial differential equation that permits a unique solution in closed form. We illustrate how to perform this construction through a specific example in Sec. 3.

The mapping $(r,\theta,\phi) \mapsto  (\alpha,\gamma,\phi)$ is not unique. It depends not only on the particular magnetic field being investigated but also on one's choice of $\gamma$ and is often cumbersome to write down. While the choice of $\alpha$ is obviously dictated by the poloidal magnetic field, the actual form of $\alpha$ can be complicated when one deals with a linear combination of many multipoles. For the relatively simple case of a dipole or dipole-plus-quadrupole field, we can choose (as we do in Sec. 3) a simple form for $\gamma$, namely $\gamma=\cos\theta$, and equation (12) simplifies. However, when $g(\theta)$ [the angular part of $\alpha(r,\theta)$] is a combination of many powers of $\sin\theta$ and $\cos\theta$, it becomes difficult to choose $\gamma$ such that equation (12) is both well-defined and simple.

\section{Worked example: dipole-quadrupole, poloidal-toroidal field}

In this section, we calculate $\delta\rho$ due to a mixed dipole and quadrupole magnetic field, with a toroidal component confined to the region bounded by the last poloidal field line that closes inside the star. The external field, to which the internal field is matched at $r=1$, is given by \citep{mlm13}

\be {\bf{B}}_\mathrm{ext}=B_0\left\{\left[\frac{2\cos\theta}{r^3}+\frac{\kappa(3\cos^2\theta -1)}{r^4}\right]\er - \left(\frac{\sin\theta}{r^3} + \frac{2\kappa\sin\theta\cos\theta}{r^4}\right)\etht\right\},\ee
where $B_0$ sets the strength of the field overall and $\kappa$ is the dimensionless weight of the quadrupole component.

The deformation caused by a field without an internal toroidal component matched to equation (13) has been calculated previously \citep{mlm13}. Therefore, in this section, we focus on the deformation caused by the toroidal component. We begin by discussing the conditions that $\alpha$ must obey, then we derive a suitable (but not unique) form of $\alpha$, i.e., one which generates an internal poloidal-toroidal magnetic field that matches ${\bf{B}}_\mathrm{ext}$ at $r=1$, for which the Lorentz force is continuous everywhere inside the star. Lastly, we calculate $\delta\rho_t$ and hence $\delta\rho$ overall.

Throughout the rest of this paper, to simplify analysis and to facilitate comparison with previous work, we use the gravitational potential, density, and pressure profiles used by \citet{metal11}, namely

\be \rho = \rho_\textrm{c}(1-r^2),\ee
\be p = p_\textrm{c}\left(1-\frac{5r^2}{2}+2r^4-\frac{r^6}{2}\right),\ee
\be \frac{\textrm{d}\Phi}{\textrm{d}r}=\frac{GM_*}{2R_*^2}r(5-3r^2),\ee
where $\rho_\textrm{c}=15M_*/(8\pi R_*^3)$ and $p_\textrm{c}=15GM_*^2/(16\pi R_*^4)$ are the density and pressure at the origin, respectively. This simple, `parabolic' density profile has been shown to result in values of $\epsilon$ that are $\sim 5\%$ of those found using the more realistic polytropic equation of state \citep{metal11}.

\subsection{Field-aligned coordinates}

The main problem with a magnetic field that is not symmetric around some $\theta$ is making sure that $\delta\rho_t$ is continuous across $\alpha=\alphac$ \citep{mlm13}. An example showing why this does not occur automatically is worked out in the Appendix. As explained in Sec. 2.2, the issue can be circumvented by working in the field-aligned coordinates $(\alpha,\gamma,\phi)$, where it is easier to choose $\alpha$ judiciously by inspection.

Consider a stream function $\alpha$ that is a linear combination of dipole and quadrupole components, viz.

\be \alpha(r,\theta)= f_1(r)\sin^2\theta + \kappa f_2(r)\sin^2\theta\cos\theta.\ee
Field and current continuity at the surface of the star require that the radially dependent factors satisfy \citep{metal11,mlm13}

\be f_{1}(1) = 1 = f_2(1),\ee
\be 2f'_1(1) = -2 = f'_2(1),\ee
\be f''_1(1) - 2 f_1(1) = 0 = f''_2(1) - 6 f_2(1),\ee
where the prime indicates differentiation with respect to $r$. The first and second conditions ensure the continuity of $B_r$ and $B_\theta$ respectively. The third condition ensures current continuity (i.e., the current vanishes at the surface). Implicitly, the current is also well-behaved at the origin.

The last closed internal poloidal field line is defined by $\alpha=\alphac$, where $\alphac$ is the maximum of $f_1(1)\sin^2\theta+\kappa f_2(1)\sin^2\theta\cos\theta$ for $0\leqslant\theta\leqslant \pi/2$. In the case of a pure dipole, there is only one neutral curve at $\theta=\pi/2$. In the case of a pure quadrupole, there are two neutral curves at $\theta=\cos^{-1}(\pm\sqrt{1/3})$. Mixed dipole-quadrupole cases may have one or two neutral curves, depending on the relative weights of the multipoles; see Fig. \ref{field} of this paper and Fig. 3 of \citet{mlm13} for some examples. For consistency and simplicity, we assume in this paper that the toroidal component only exists around the upper hemisphere's neutral curve.

Now, suppose we take $\gamma = \cos\theta$. This choice is not unique, but it has the advantages that it is single-valued in the domain $0\leqslant\theta\leqslant\pi$ and that, trivially, we have $\partial\gamma/\partial r=0$. Next, we need to find suitable radial functions $f_1(r)$ and $f_2(r)$ that satisfy the above conditions (18)--(20), while bearing in mind that equation (12) is a well-defined linear partial differential equation in $\alpha$ and $\gamma$, i.e., there are no `stray' factors of $r$, $\sin\theta$, or $\cos\theta$ left. We choose $f_1(r)$ and $f_2(r)$ to be polynomials in $r$, which are simple and analytically tractable. We need at least three terms in each polynomial, so as to satisfy the three boundary conditions above. The implicit condition on the current at the origin is easily satisfied by choosing the starting powers of $r$ high enough.

\begin{figure}
\centerline{\epsfxsize=14cm\epsfbox{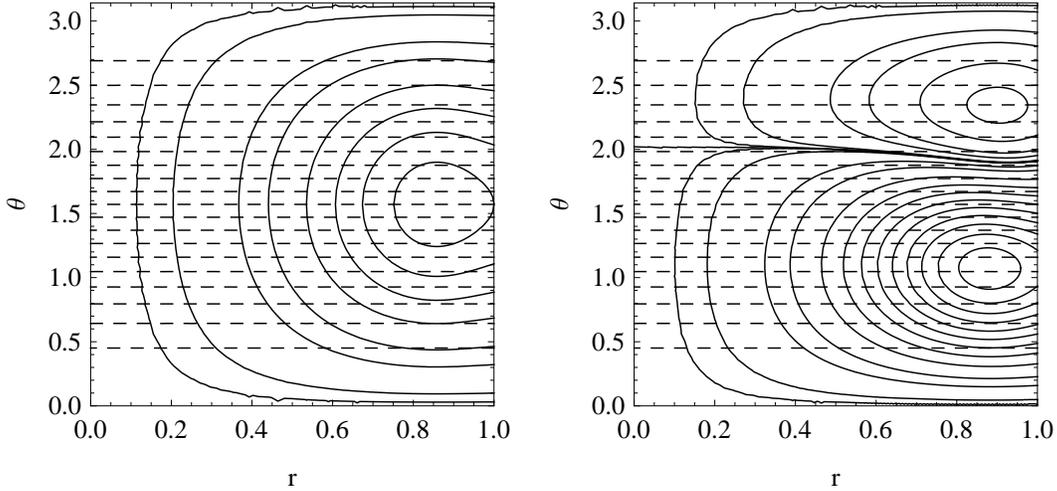}}
 \caption{Contour plots of field-aligned coordinates $\alpha(r,\theta)$ (solid curves) and $\gamma(r,\theta)$ (dashed curves) on the $r$--$\theta$ plane [from equations (17), (21), and (22)] for a pure dipole ($\kappa=0$; left panel) and a mixed dipole-quadrupole ($\kappa=2$; right panel) with $\sigma=-2$ fixed.}
 \label{coordp}
\end{figure}

\begin{figure}
\centerline{\epsfxsize=14cm\epsfbox{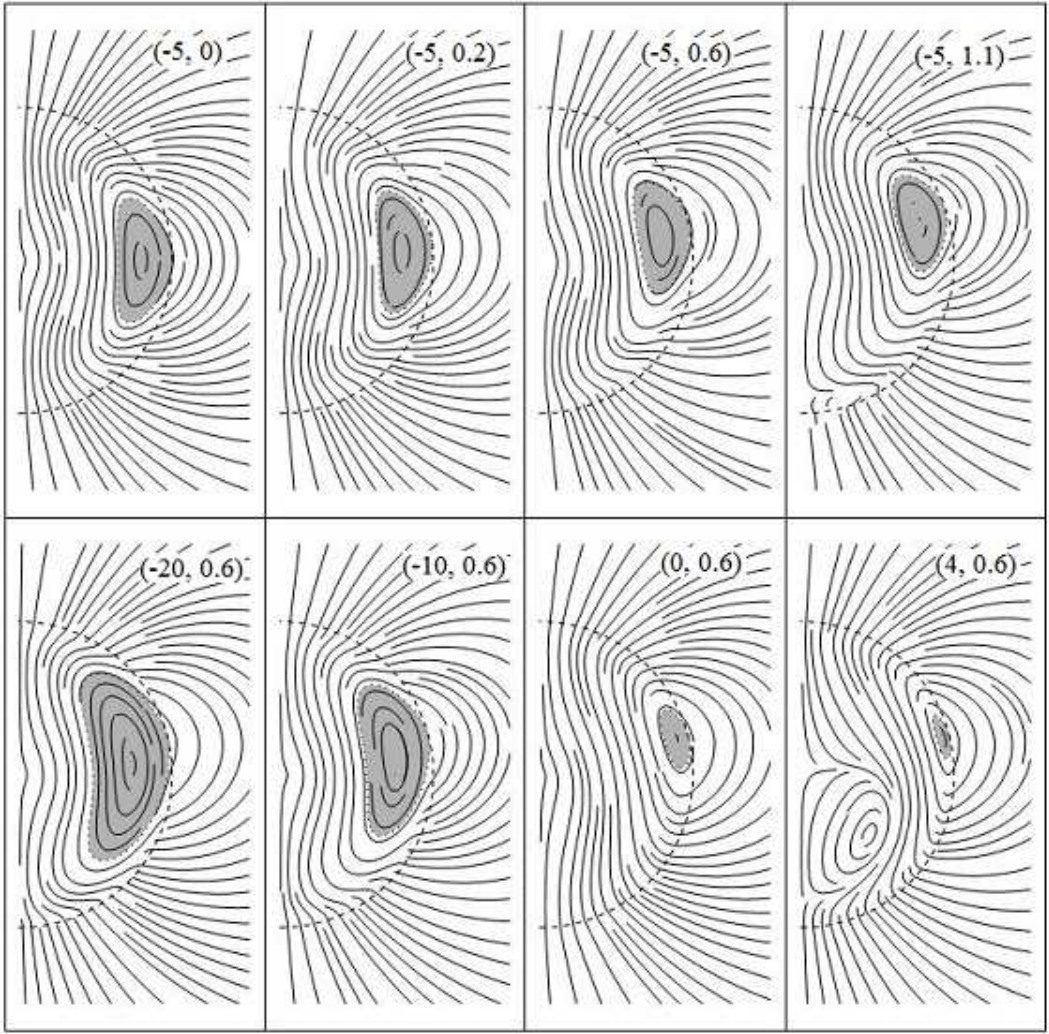}}
 \caption{Magnetic field lines of a mixed dipole-quadrupole configuration, as a function of toroidal volume (related to $\sigma$) and quadrupole coefficient $\kappa$.  The internal field is described by equations (1), (2), and (17)--(22). The external field is described by equation (13). Each panel is labelled by ($\sigma$, $\kappa$). Top panels: $\sigma= 0.5$; $\kappa=0$, 0.2, 0.6, and 1.1 (left to right). Bottom panels: $\kappa= 0.6$; $\sigma=-20$, $-10$, 0, and 4 (left to right). In each panel, the shading defines the volume to which the toroidal magnetic field component is confined. The dashed semicircle represents the stellar surface.}
 \label{field}
\end{figure}

In fact, even after applying these constraints, there are still an infinite number of polynomials that meet our needs. We make the final choice through trial and error, to ensure that $r(\alpha,\gamma)$ is real, with $0\leqslant r(\alpha,\gamma)\leqslant 1$ in the $(\alpha,\gamma)$ domain of interest, and that it is expressible in closed form (for ease of calculation). We note that $a_1 r^{n}+a_2 r^{2n}+a_3 r^{3n}$ is a cubic in $r^{n}$ and is guaranteed to have at least one real root. We note also that the lowest power of $r$ for which the quadrupole's current vanishes at the origin is $n=3$, but $f_1(r)$ cannot have an $r^3$ term, if the dipole's current is to vanish at the origin, so one needs $n\geqslant 4$. In addition, we add an arbitrary, extra term to the $f_1(r)$ polynomial, whose coefficient $\sigma$ is not fixed by any of the boundary conditions; it is left as a free parameter. Adjusting $\sigma$ changes the volume of the toroidal field region, a quantity of key physical importance, in a convenient fashion \citep{mlm13,detal14}. Having chosen a value of $\sigma$, one can then check for the existence of real roots in the domain $0\leqslant r\leqslant 1$; the existence is guaranteed for $\sigma=0$ and for the values of $\sigma$ discussed in this paper. Now, choosing the lowest power of $r$ in both $f_1(r)$ and $f_2(r)$ to be $r^4$, we finally arrive at

\be f_1(r)=\left(\frac{117}{32}-\sigma\right)r^4 - \left(\frac{65}{16}-3\sigma\right)r^8 + \left(\frac{45}{32}-3\sigma\right)r^{12}+\sigma r^{16},\ee
\be f_2(r)=\frac{1}{8}\left(35 r^4 - 42 r^8 + 15 r^{12}\right).\ee
The choices leading to equations (21)--(22) are not unique, but they do guarantee a well-defined inverse coordinate transform relation $r(\alpha,\gamma)$ for all $\kappa$ by inverting $\alpha=f_1(r)(1-\gamma^2) + \kappa f_2(r) \gamma (1-\gamma^2)$. We plot two examples of $\alpha(r,\theta)$ and $\gamma(\theta)$ on the $r$--$\theta$ plane in Fig. \ref{coordp} for $\kappa=0$ (i.e., a pure dipole, left-hand panel) and $\kappa=2$ (right-hand panel) with $\sigma=-2$ fixed. Contours of $\alpha$ follow the poloidal field lines. The pure dipole case is symmetric around the equator, but the mixed case is not. As the right-hand panel shows, a mixed dipole-quadrupole case can have two neutral curves, one in each hemisphere.

Eight example field-line plots with constant $\sigma$ and varying $\kappa$ (top four panels), and constant $\kappa$ and varying $\sigma$ (bottom four panels) are shown in Fig. \ref{field}. The top leftmost panel shows the $\kappa=0$ configuration (i.e., a pure dipole). The other top panels show configurations with $\kappa=0.2$, 0.6, and 1.1. The three $\kappa\neq 0$ plots show that a quadrupole component breaks the north-south symmetry of the field lines. Odd multipoles are north-south antisymmetric, while even multipoles are north-south symmetric; when odd and even multipoles are added together, the result is generally asymmetric \citep{mlm13}. The bottom panels show a configuration with $\kappa=0.6$ and different sizes of the toroidal region. As $\sigma$ increases, the toroidal field volume decreases. A pure dipole has one neutral curve at $\theta=\pi/2$, but the addition of a quadrupole shifts this original neutral curve northwards and introduces a new neutral curve into the southern hemisphere. As the bottom right panel of Fig. \ref{field} shows, the new neutral curve also grows as the original toroidal region shrinks, even when $\kappa$ is kept constant. The general asymmetry of these composite fields, both internal and external, can potentially be linked to observables such as asymmetry in neutron star emission and burst signals (see Sec. 4).

The fields in Fig. \ref{field}, generated by equations (17), (21), and (22), have smaller magnitudes at the origin than those analysed previously \citep{metal11,aetal13,mlm13,detal14}, as indicated by the sparseness of field lines around the origin. Indeed, some field lines seem to bend away from the origin, as in Fig. \ref{field}. This is a consequence of the higher powers of $r$ in $f_1(r)$ and $f_2(r)$ used in this paper compared to previous works. The detailed field behaviour at $r=0$ is not a major concern here, as our primary focus is on stellar ellipticity, and the region near the origin contributes little to the moment of inertia.

\subsection{Deformation}

The density perturbations due to the toroidal and poloidal magnetic components, $\delta\rho_t$ and $\delta\rho_p$, are readily calculated from equations (10) and (11) respectively. The total density perturbation, $\delta\rho_p+\delta\rho_t$, is then substituted into equations (4) and (5) to calculate $\epsilon$.

\begin{figure}
\centerline{\epsfxsize=14cm\epsfbox{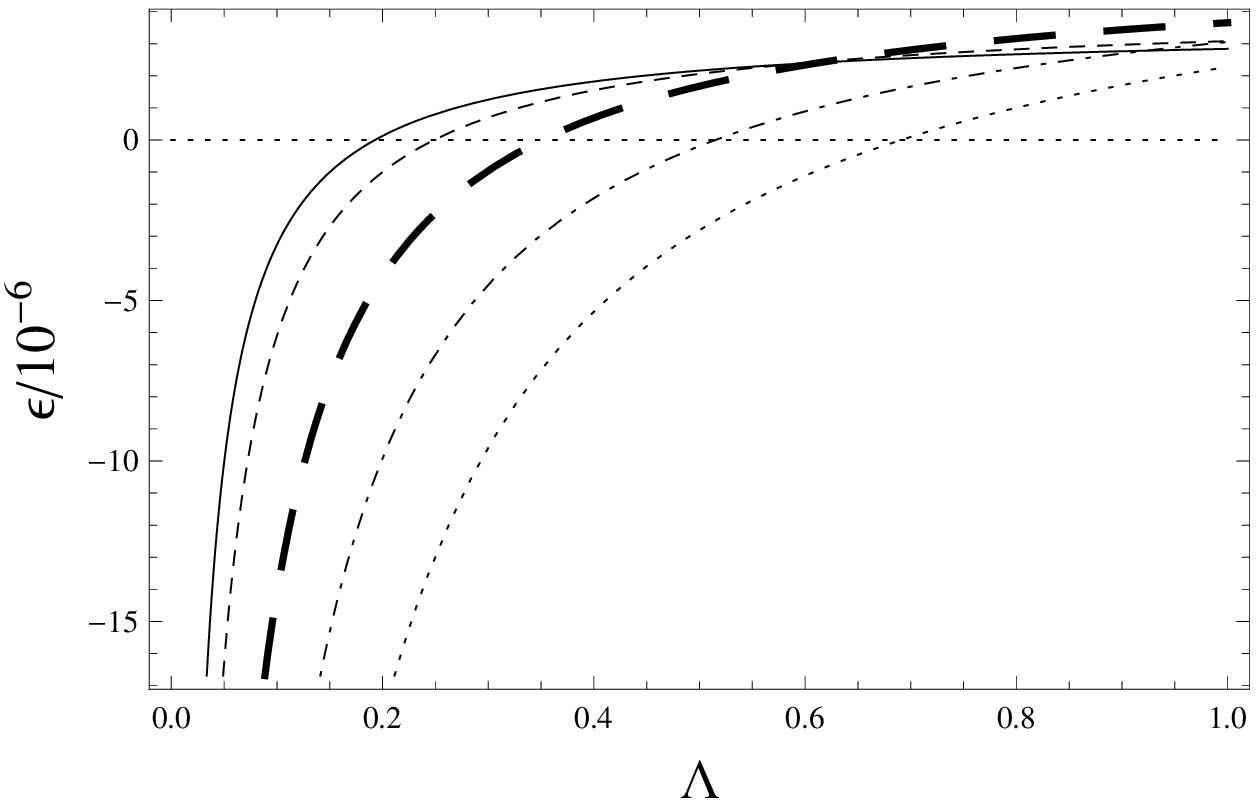}}
 \caption{Mass ellipticity $\epsilon$ (in units of $10^{-6}$) versus $\Lambda$ (the ratio of internal poloidal magnetic field energy to total internal magnetic field energy), for $\sigma=-5$ fixed and $\kappa= 0$ (thin solid curve), 0.2 (thin dashed curve), 0.6 (thin dashed-dotted curve), 1.1 (thin dotted curve), for a canonical magnetar ($M=1.4$ $\msun$, $\rstar=10^4$ m, $B_0=5\times 10^{10}$ T). The result for a pure dipole \citep{metal11} is included for comparison (thick dashed curve).}
 \label{evsl}
\end{figure}

\begin{figure}
\centerline{\epsfxsize=14cm\epsfbox{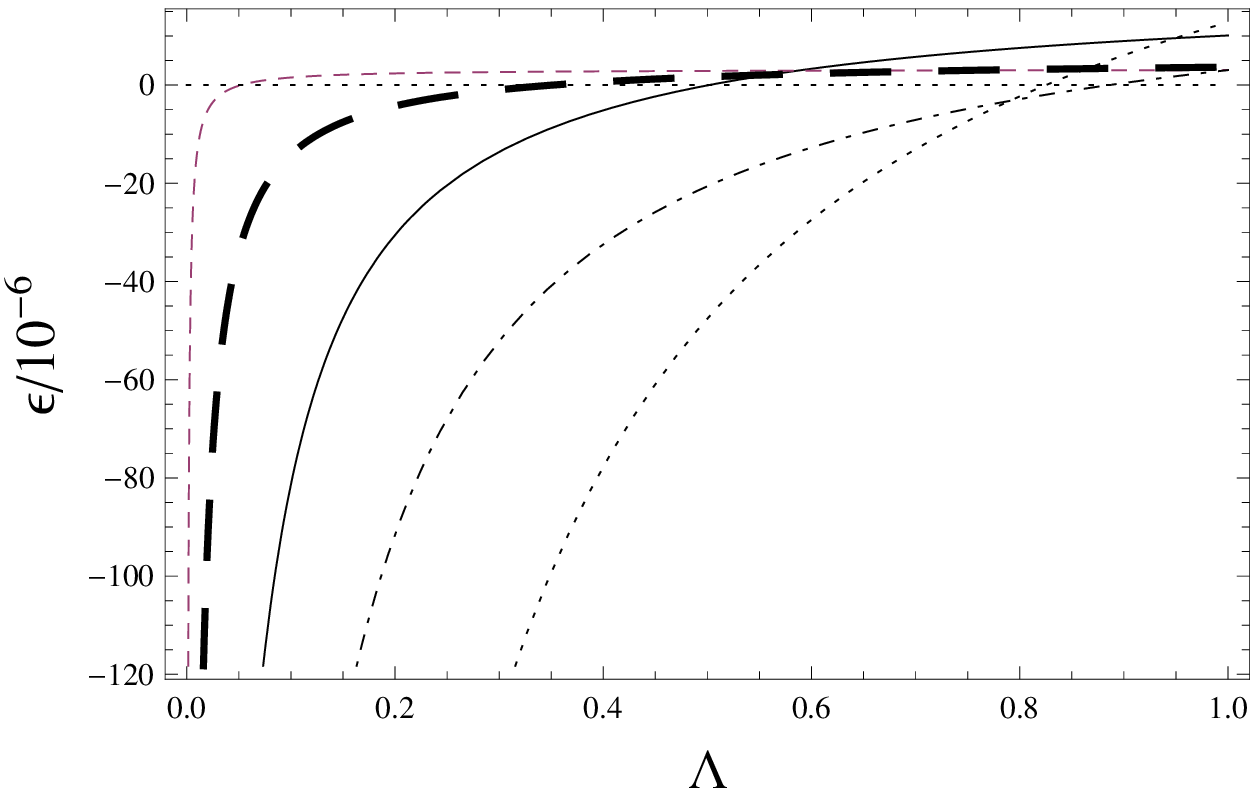}}
 \caption{Mass ellipticity $\epsilon$ (in units of $10^{-6}$) versus $\Lambda$ (the ratio of internal poloidal magnetic field energy to total internal magnetic field energy), for $\kappa=0.2$ fixed and $\sigma= -15$ (thin solid curve), $-10$ (thin dashed curve), 10 (thin dashed-dotted curve), 15 (thin dotted curve, for a canonical magnetar ($M=1.4$ $\msun$, $\rstar=10^4$ m, $B_0=5\times 10^{10}$ T). The result for a pure dipole \citep{metal11} is included for comparison (thick dashed curve).}
 \label{evsl2}
\end{figure}

In Fig. \ref{evsl}, we plot $\epsilon$ as a function of the parameter $\Lambda$ (the ratio of internal poloidal field energy to total internal field energy; $0\leqslant\Lambda\leqslant 1$) for a canonical magnetar ($M=1.4$ $\msun$, $\rstar=10^4$ m, $B_0=5\times 10^{10}$ T) for fixed $\sigma$ and four different values of $\kappa$. In Fig. \ref{evsl2}, we plot $\epsilon$ versus $\Lambda$ for fixed $\kappa$ and four different values of $\sigma$. In both figures, we also plot $\epsilon(\Lambda)$ for a purely dipolar field with the same $M$, $\rstar$, and $B_0$ as the thick dashed curve \citep{metal11}. The plots show that prolateness increases as $\Lambda$ decreases (i.e., as the toroidal magnetic field energy increases), similar to the results of \citet{metal11} and \citet{detal14}. Figure \ref{evsl} shows that, for a given $\sigma$, the curve shifts rightward and downward as the quadrupole field component strengthens, i.e., the star tends to be more prolate for the same $\Lambda$ as $\kappa$ increases. In contrast, a star with a purely \emph{poloidal} magnetic field becomes more oblate, as the quadrupole component strengthens \citep{mlm13}. The curves intersect at some points. For example, we find $\epsilon(\sigma=-5,\kappa=0)=\epsilon(\sigma=-5,\kappa=0.2)$ at $\Lambda\approx 0.6$. If one were to detect a magnetar near these particular values of $\epsilon$ and $\Lambda$, the analysis in this paper cannot distinguish between the two models.

The curves in Fig. \ref{evsl2} show the same general trend: prolatenes increases as $\Lambda$ decreases. The curves for $\sigma=-10$ (dashed) and $\sigma=-15$ (solid) indicate that the star becomes more prolate, as the region occupied by the toroidal field grows. However, the curves for $\sigma=10$ (dashed-dotted) and $\sigma=15$ (dotted) indicate that the star becomes more prolate, as the toroidal field region shrinks below some volume (for this particular example, the star becomes more prolate again for $\sigma\gtrsim 2$). This counterintuitive trend occurs, because a second neutral curve (and the corresponding region with low poloidal field strength) emerges inside the star in configurations with high and positive $\sigma$. The weak poloidal field around the second neutral curve makes it easier to deform the star into a prolate shape. The curves intersect at some points, like in Fig. \ref{evsl}. For example, we find $\epsilon(\sigma=-15,\kappa=0.2)=\epsilon(\sigma=-10,\kappa=0.2)$ at $\Lambda=0.58$, $\epsilon(\sigma=10,\kappa=0.2)=\epsilon(\sigma=15,\kappa=0.2)$ at $\Lambda=0.78$, and $\epsilon(\sigma=-10,\kappa=0.2)=\epsilon(\sigma=15,\kappa=0.2)$ at $\Lambda=0.86$.

\citet{y13} calculated $\epsilon$ for a pure dipole without taking the Cowling approximation. His values of $|\epsilon|$ are $\sim 2$ times those found by \citet{metal11}. Furthermore, \citet{y13} found that, if the zeroth-order density profile $\rho(r)$ features an off-centred maximum (a somewhat artificial situation), taking the Cowling approximation results in the opposite shape from that predicted by the full perturbation analysis (i.e., a prolate star where the full perturbation analysis predicts an oblate star and vice versa). A thorough, full-perturbation calculation without the Cowling approximation is beyond the scope of this paper but is certainly worth undertaking in future.

\subsection{Magnetic moment of inertia}

In Sec. 3.2, we neglect the direct contribution to $\epsilon$ from the magnetic energy density in the $T^{00}$ component of the stress-energy tensor.\footnote{We thank the reviewer for bringing this issue to our attention.} Most recent works on this subject [e.g., \citet{hetal08,detal14}] neglect this contribution as well. It is given by \citep{t80}

\be \epsilon_B = \frac{\pi\rstar^5}{I_0}\int_0^{\pi}\int_0^1 \textrm{d}r \phantom{i}\textrm{d}\theta\phantom{i} r^4\sin\theta(1-3\cos^2\theta)\frac{B^2}{2\mu_0 c^2},\ee
which is essentially the same as equation (13) of \citet{metal11}, with $\delta\rho$ replaced by $B^2/2\mu_0 c^2$. The external field contribution to $\epsilon_B$ is comparable to equation (23), because externally we have $B_\textrm{ext}^2\propto r^{-6}$ [equation (13)] and $\int_{R_\ast}^\infty \textrm{d}r\phantom{i} r^4B^2$ is dominated by the lower terminal. For the dipole-plus-quadrupole example in Sec. 3.2 with $\sigma=-5$ and $\kappa=1.1$ (corresponding to the top rightmost panel of Fig. 2 and the thin dotted curve in Fig. 3), we find

\be \epsilon_B = -1.4\times 10^{-7} \left(\frac{B_0}{5\times 10^{10}\textrm{ T}}\right)^2 \left(\frac{\mstar}{1.4\phantom{i}\msun}\right)^{-1} \left(\frac{\rstar}{10^4 \textrm{ m}}\right)^3 \left(1-\frac{0.847}{\Lambda}\right).\ee
Note that $\epsilon_B$ scales differently with $\rstar$ and $\mstar$ [compared to the $\delta\rho$ contribution to $\epsilon$, e.g., given by equation (25) of this paper or equation (17) of \citet{metal11}], because $\epsilon_B$ is not coupled to $\nabla\Phi$ [given by equation (16) in this paper for the parabolic density profile].

We plot $\epsilon$ versus $\Lambda$ for a canonical magnetar ($M=1.4$ $\msun$, $\rstar=10^4$ m, $B_0=5\times 10^{10}$ T), with $\sigma=-5$ and $\kappa=1.1$ in Fig. 5. The figure shows two nearly overlapping curves. The dashed (solid) curve corresponds to excluding (including) the magnetic field contribution. Our calculation confirms that, at worst, the effect shifts $\epsilon$ by a factor of $\sim 2\%$. This is because one has $\delta\rho/(B^2/\mu_0 c^2)\sim c^2(\textrm{d}r/\textrm{d}\Phi)\gg 1$.

The focus of this methods paper is on the star-deforming effect of the magnetic field. The uncertainty introduced by neglecting $\epsilon_B$ is smaller than that potentially introduced by the commonly-taken Cowling approximation \citep{y13}, nor does it change the general qualitative behaviour of the results.

\begin{figure}
\centerline{\epsfxsize=14cm\epsfbox{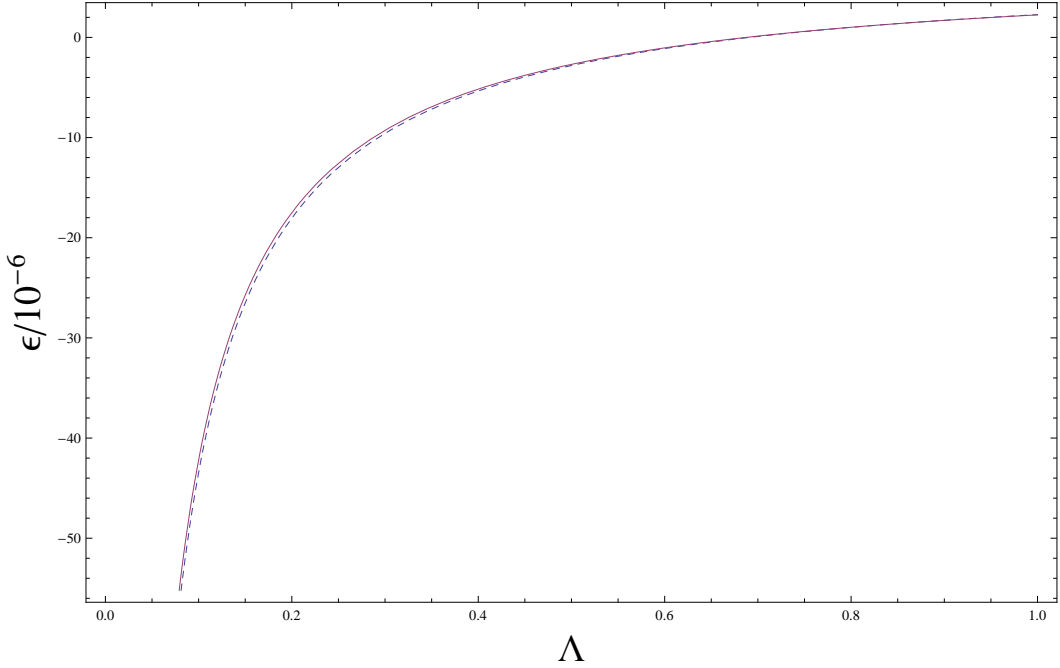}}
 \caption{Mass ellipticity $\epsilon$ (in units of $10^{-6}$) versus $\Lambda$ (the ratio of internal poloidal magnetic field energy to total internal magnetic field energy), for a dipole-plus-quadrupole field configuration, with $\sigma=-5$ and $\kappa= 1.1$, where we take into account (solid curve) and neglect (dashed curve) the moment-of-inertia tensor contribution of the magnetic field, for a canonical magnetar ($M=1.4$ $\msun$, $\rstar=10^4$ m, $B_0=5\times 10^{10}$ T). The two curves nearly overlap.}
 \label{ellmag}
\end{figure}

\section{Astrophysical examples}

While this paper is mainly a `methods' paper, we discuss briefly, by way of illustration, two examples of how one can apply the method to model astrophysical phenomena in the magnetars SGR 0418+5729 \citep{gog08,ggo11,tetal13} and 4U 0142+61 \citep{metal14}.

\subsection{SGR 0418+5729}

SGR 0418+5729 has an inferred dipole field strength of $\lesssim 7.6\times 10^8$ T \citep{retal10}, but an analysis of the X-ray spectrum by \citet{ggo11} concluded that a surface field strength of $10^{10}$ T fits the data best. \citet{ggo11} suggested that the surface field contains higher-order multipole(s) (which fall away with $r$ faster than the dipole) to account for the discrepancy. If we take $B_0=3.8\times 10^8$ T, corresponding to a dipole field strength of $7.6\times 10^8$ T at the polar surface, and take $\kappa=13$, corresponding to \emph{total} (dipole plus quadrupole) field strength of $10^{10}$ T at the polar surface, then we find

\be \epsilon = 1.7\times 10^{-1} \left(\frac{B_0}{5\times 10^{10} \textrm{ T}}\right)^2 \left(\frac{\mstar}{1.4\msun}\right)^{-2} \left(\frac{\rstar}{10 \textrm{ km}}\right)^4\left(\frac{\Lambda-1+7\times 10^{-4}}{\Lambda}\right),\ee
for $\sigma =0$. We choose to write $\epsilon$ in terms of $B_0$ (which can, in principle, be inferred from spin and spin down) rather than (say) the internal magnetic field strength (maximum or volume-averaged) in order to relate $\epsilon$ to an observable quantity, which does not itself depend on $\Lambda$. Furthermore, this facilitates comparison with other works \citep{metal11,detal14}.

The above model does not need much toroidal component to deform into a prolate shape; we obtain $\epsilon\leqslant 0$ for $1-\Lambda\geqslant 7\times 10^{-4}$. If SGR 0418+5729 has no toroidal field component, we obtain $\epsilon = 6.8\times 10^{-9}(\mstar/1.4\msun)^{-2}(\rstar/10\textrm{ km})^4$, which is too small to generate gravitational waves detectable by current-generation interferometers \citep{metal11,r13}. However, $\Lambda=0.9$ is enough to result in $\epsilon=- 10^{-6}(\mstar/1.4\msun)^{-2}(\rstar/10\textrm{ km})^4$, in contrast with a purely dipolar neutron star, which only becomes significantly prolate for $\Lambda\ll 0.4$ \citep{metal11}. Thus, in objects like SGR 0418+5729, $\epsilon$ depends strongly on field \emph{configuration}, characterized by $\Lambda$, as well as field magnitude. Future gravitational-wave upper limits from next-generation detectors will constrain $\Lambda$ better in the context of this model. Because of the low spin frequencies of magnetars, e.g., 0.11 Hz for SGR 0418+5729 \citep{gwk09}, the best limits will come from the Einstein Telescope, where Newtonian noise suppression will extend observations down to the sub-Hertz regime \citep{bvem10}. Present-day limits on $\epsilon$ complement historical limits based on fast-spinning birth scenarios \citep{tcq04,ds07,dss09,mp14}.

\subsection{4U 0142+61}

The magnetar 4U 0142+61 has an inferred dipole field strength of $1.3\times 10^{10}$ T \citep{gk02,gog08}. Phase modulations of its hard X-ray pulses, observed by the Suzaku satellite \citep{eetal11,metal14}, have been interpreted by \citet{metal14} as evidence of free precession, indicating a prolate star with $|\epsilon|\approx 1.6\times 10^{-4}$ and, hence, a maximum internal toroidal field $B_t\sim 10^{12}$ T.

As in the previous subsection, we assume that the inferred dipole field is not the whole story: the actual field contains a quadrupole element, which cannot be inferred from spin down, and an internal toroidal field, whose presence can only be detected indirectly via its effect on $\epsilon$. From equation (13), the polar surface field strengths of the dipole and quadrupole components are $B_\mathrm{dip}=2B_0$ and $B_\mathrm{quad}=2\kappa B_0$ respectively. If we infer that $B_\mathrm{dip}=1.3\times 10^{10}$ T from spin down, we can calculate the values of $\Lambda$ required to obtain $\epsilon=-1.6\times 10^{-4}$, as required by the precession interpretation, for different values of $\kappa$ (with $\sigma=-0.1$ in all cases). We present the cases of $\kappa=0.78,$ 1, 2, 5, and 10 in Table 1. Each row shows $\kappa$, $B_\mathrm{pole}$, which is the total polar surface field strength, $\Lambda_{\epsilon}$, which is the value of $\Lambda$ required to impart $\epsilon=-1.6\times 10^{-4}$ to the star, and $B_t$, which is the maximum strength of the toroidal field. Note that $\kappa=0.78$ is chosen because this corresponds to $B_\mathrm{pole}=4.7\times 10^{10}$ T, the value obtained by \citet{gog08} when they analysed the X-ray pulses from 4U 0142+61.

For the cases investigated, we find values of $B_t$ that are $\sim 10$ times stronger than calculated by \citet{metal14}. This, however, can be counteracted by reducing $\sigma$ and, hence, enlarging the volume occupied by the toroidal field. Table 1 shows that, as the quadrupole component strengthens, $\Lambda_\epsilon$ increases, i.e., a more quadrupolar field needs less toroidal field energy to deform the star. However, it is harder to say with confidence what trend $B_t$ follows, e.g., $B_t$ decreases between $\kappa=0.78$ and 2, then increases between $\kappa=2$ and 5, then decreases again between $\kappa=5$ and 10. This is because $\kappa$ affects the toroidal field's volume as well, although to a lesser degree than $\sigma$. Thus, for example, we find $\Lambda_\epsilon(\kappa=2)<\Lambda_\epsilon(\kappa=5)$, i.e., the $\kappa=5$ case requires less toroidal field energy to deform the star to the desired $\epsilon$. However, we also find $B_t(\kappa=2)<B_t(\kappa=5)$, because the toroidal field's volume for $\kappa=5$ is smaller than that for $\kappa=2$.

Note that we do not claim the magnetar 4U 0142+61 to be in any of the magnetic configurations explored in this subsection. We present these examples simply to demonstrate what our analytical approach can accommodate and analyse.

\begin{table*}

 \begin{minipage}{145mm}
 \centering
  \caption{Sample models of 4U 0142+61 for $\epsilon=-1.6\times 10^{-4}$, the ellipticity required by interpreting X-ray pulse modulation as precession \citep{metal14}, for some different values of $\kappa$, with $\sigma=-0.1$ fixed. We show $\kappa$ (first column), total polar surface field strength $B_\mathrm{pole}$ (second column), the required poloidal-to-total field energy ratio $\Lambda_\epsilon$ (third column), and the maximum toroidal field strength $B_t$ (fourth column).}
  \begin{tabular}{@{}lccc@{}}
  \hline
    $\kappa$ & $B_\mathrm{pole}$ ($10^{12}$ T) & $\Lambda_\epsilon$ & $B_t$ ($10^{12}$ T)\\
\hline
0.78 &  $4.7\times 10^{-2}$ & $2.8\times 10^{-4}$ & $16$\\
1 &  $5.2\times 10^{-2}$ & $9.2\times 10^{-4}$ & $9.5$\\
2 &  $7.8\times 10^{-2}$ & $2.7\times 10^{-3}$ & $7.1$\\
5 &  $1.6\times 10^{-1}$ & $6.2\times 10^{-3}$ & $9.3$\\
10 &  $2.9\times 10^{-1}$ & $3.8\times 10^{-2}$ & $6.9$\\
\hline
\end{tabular}
\end{minipage}
\end{table*}

\section{Conclusion}

In this short methods paper, we generalise previous calculations \citep{metal11,mlm13} to show how any multipolar stellar magnetic field containing both poloidal and toroidal axisymmetric components can be constructed analytically to satisfy boundary conditions (zero surface currents and zero toroidal field outside the equatorial torus) motivated physically and by numerical simulations in the literature. We also present a worked example of a dipole-plus-quadrupole, poloidal-plus-toroidal field and calculate $\epsilon$ versus $\Lambda$ for some representative combinations of the parameters $\sigma$ (which controls the volume occupied by the toroidal field) and $\kappa$ (which controls the weight of the quadrupole component). The star tends to be more prolate as $\kappa$ increases (Fig. \ref{evsl}). In the fixed $\kappa=0.2$ example shown in Fig. \ref{evsl2}, for $\sigma\lesssim 2$, the star deforms into a more prolate shape as $\sigma$ decreases. For $\sigma\gtrsim 2$, increasing $\sigma$ makes the star more prolate, as the internal poloidal field component weakens. For a given $\Lambda$, i.e., for a given toroidal field energy, smaller $\sigma$ means greater toroidal field strength.

In Sec. 4, we briefly discuss two possible astrophysical applications of our analysis to magnetars. SGR 0418+5729 is interesting because there is a mismatch between the surface field strengths inferred from spin down \citep{retal10} and from X-ray analysis \citep{ggo11}. We show how, if the star's magnetic field is mostly quadrupolar [as suggested by \citet{ggo11}], the star is likely to be prolate. For $\kappa=13$, we find $1-\Lambda\geqslant 7\times 10^{-4}$ for $\epsilon \leqslant 0$ (i.e., the toroidal field only needs to contribute at least $0.07\%$ to the total field energy to deform the star into a prolate shape). Present-day and historical (at birth) upper limits on gravitational wave emissions can thus be used to infer the internal field configuration and toroidal field strength. 4U 0142+61 is interesting, because Suzaku observations suggest that it undergoes free precession, from which an upper limit for $\epsilon$ can be calculated \citep{metal14}. We present several possible configurations of this magnetar's field and calculate the lower limit on $\Lambda$ implied by the upper limit on $\epsilon$. We find that, to obtain $|\epsilon|\sim 10^{-4}$, we need toroidal fields with maximum strength $B_t\sim 10^{13}$ T, ten times greater than inferred by \citet{metal14}. The two applications are by no means exhaustive; they simply illustrate the potential of the method.

Throughout this paper, we do not discuss the stability of these dipole-plus-quadrupole, poloidal-plus-toroidal field configurations. Such calculations are reserved for future work. The results of \citet{aetal13} are not directly applicable to the composite multipolar fields discussed here. To repeat their analysis, one must first identify the regions of greatest instability in the star. This is more difficult than for the pure dipole field, because the field is no longer north-south symmetric, because the toroidal field is offset from the equator, and because there are new regions of low field strength inside the star (that is, other than the neutral curve). It may be easier to test the stability of these composite fields numerically rather than analytically, by evolving the field in a time-dependent magnetohydrodynamic simulation. We defer this calculation to a future paper.

\appendix

\section{An example of a discontinuous $\delta\rho_{\MakeLowercase{t}}$ arising from a poor choice of $\MakeLowercase{f(r)}$}

In this short Appendix, we present an example of what happens when one uses the simple form of $\beta(\alpha)$ given in equation (7) without choosing the polynomial $f(r)$ judiciously. We start with a pure quadrupole field,

\be \alpha = f_2(r)\sin^2\theta\cos\theta,\ee
\be \beta(\alpha) =
\begin{cases}
(|\alpha| - \alpha_c)^2&\textrm{for }|\alpha|\geqslant \alpha_c,\\
0&\textrm{for }|\alpha| < \alpha_c,
\end{cases}
\ee
with $\alpha_c=2\sqrt{3}/9$. In this section, we choose the radial function $f_2(r)=21(r^3 - \frac{5}{3} r^4 + \frac{5}{7} r^5)$, as chosen by \citet{mlm13}. This $f(r)$ is a natural choice, because it fulfills the boundary conditions (i)--(v) listed in Sec. 2.1, because it has the minimum number of terms, and because its first power, $r^3$, is the lowest power of $r$ that guarantees the current vanishes at the origin. However, this choice of $f(r)$ does not allow us to perform an invertible coordinate transformation $(r,\theta,\phi)\mapsto (\alpha,\gamma,\phi)$, and therefore does not permit a unique solution in closed form for equation (12).

\begin{figure}
\centerline{\epsfxsize=18cm\epsfbox{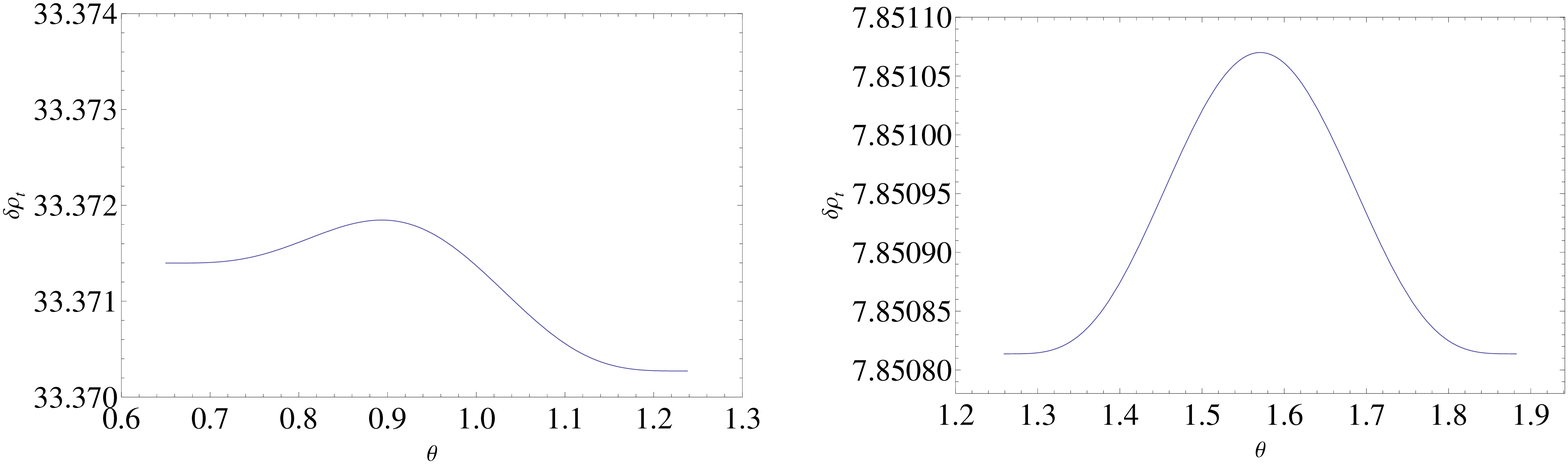}}
 \caption{Density perturbation due to the toroidal field component, $\delta\rho_t$, versus colatitude $\theta$ for $r=0.7$, for a quadrupole where the field is described by equations (A1)--(A2) (left-hand panel) and for a pure dipole (right-hand panel).}
 \label{app2}
\end{figure}

Substitution of equations (A1)--(A2) into equation (10) leads to

\be -\frac{\mu_0\rstar}{B_0^2}\frac{\mathrm{d}\Phi}{\mathrm{d}r}\frac{\partial\delta\rho_t}{\partial\theta}= \frac{\partial}{\partial\theta}\left[\frac{2(\alpha-\alpha_c)^3}{r^2\sin^2\theta}\frac{\partial\alpha}{\partial r}\right]-\frac{\partial}{\partial r}\left[\frac{2(\alpha-\alpha_c)^3}{r^2\sin^2\theta}\frac{\partial\alpha}{\partial\theta}\right].\ee
Integrating equation (A3) with respect to $\theta$ should give $\delta\rho_t$, plus some integrating constant $g(r)$, to be calculated by matching $\delta\rho_t=0$ at $\alpha=\alpha_c$.  We plot $\delta\rho_t$ versus $\theta$ for $r=0.7$ for the quadrupole in the left-hand panel of Fig. \ref{app2} as an example. We obtain $\delta\rho_t(r=0.7, \theta=\theta_1)\neq\delta\rho_t(r=0.7, \theta=\theta_2)$, with $\theta_1$ and $\theta_2$ being the coordinates where $\alpha(r=0.7,\theta)=\alpha_c$. For comparison, the right-hand panel of Fig. \ref{app2} shows $\delta\rho_t(r=0.7)$ for a pure dipole, showing $\delta\rho_t(r=0.7, \theta=\theta_1)=\delta\rho_t(r=0.7, \theta=\theta_2)$ \citep{metal11}. It is therefore impossible to ensure the continuity of $\delta\rho_t$ at $\alpha=\alpha_c$ for a quadrupole (unlike for the dipole) using only an arbitrary function of $r$. This problem vanishes when the method described in Sections 2.2 and 3 is applied.

\bsp \label{lastpage}

\end{document}